\documentclass[AMA,STIX2COL]{ama/WileyNJD-v2}
\usepackage{moreverb}

\usepackage{multirow}
\usepackage{multicol}
\usepackage{blindtext}
\usepackage{booktabs}
\usepackage{longtable}

\articletype{IJNM PrePrint}

\begin{document}

\title{Modeling and Accomplishing the BEREC Network Neutrality Policy\protect\thanks{BEREC - Body of European Regulators for Electronic Communications.}}

\author[1]{David S. Barreto}

\author[1,2]{Rafael F. Reale}

\author[1]{Joberto S. B. Martins*}

\authormark{D. Barreto \textsc{et al}}

\address[1]{\orgdiv{PPGCOMP}, \orgname{Salvador University - UNIFACS}, \orgaddress{\state{Bahia}, \country{Brazil}}}

\address[2]{\orgname{Federal Institute of Education - IFBA}, \orgaddress{\state{Bahia}, \country{Brazil}}}

\corres{*Joberto S. B. Martins, Salvador University - UNIFACS. PPGCOMP, Campus CTN, Salvador, Bahia, Brazil\\
\email{joberto.martins@gmail.com}}

\abstract[Abstract]{Network neutrality (NN) is a principle of equal treatment of data in network infrastructures with fairness and universality being the primary outcomes of the NN management practice. For networks, the accomplishment of NN management practice is essential to deal with heterogeneous user requirements and the ever-increasing data traffic. Current tools and methods address the NN problem by detecting network neutrality violations and detecting traffic differentiation. This paper proposes the NN-PCM (Network Neutrality Policy Conformance Module) that deploys the BEREC network neutrality policy using a bandwidth allocation model (BAM). The NN-PCM new approach allocates bandwidth to network users and accomplishes the BEREC NN policy concomitantly. Network neutrality is achieved by grouping users with similar traffic requirements in classes and leveraging the bandwidth allocation model's characteristics. The conceptual analysis and simulation results indicate that NN-PCM allocates bandwidth to users and accomplishes BEREC network neutrality conformance \emph{by design} with transparent, non-discriminatory, exceptional, and proportional management practices.}


\jnlcitation{\cname{%
\author{D. S. S. Barreto}, 
\author{R. F. Reale}, and
\author{J. S. B. Martins}} (\cyear{2020}), 
\ctitle{On Modeling and Accomplishing the BEREC Network Neutrality Policy}, \cjournal{Int J of Network Mgmt}, \cvol{2020;00:1--21}.}

\maketitle

\section{Introduction}\label{sec1}

Network neutrality (NN) is a principle of equal treatment of data moving across a network infrastructure. It is a network operation, administration, and management principle in which the  traffic should not be discriminated based on its source, destination or content \cite{stocker_2020} \cite{van_schewick_network_2015} \cite{wu_network_2003}.

The current computer network evolution scenario includes networks such as IoT (Internet of Things), 5G and Cloud/Fog/Edge computing with a wide variety of services, applications, billions of users, and ever-increasing data traffic. Corporate networks, Internet Service Providers (ISPs), and Content Providers (CPs) play a crucial role in conveying data worldwide. For all these networks, the NN modeling and deployment is a challenging task not yet adequately addressed by the research community due to its heterogeneous traffic characteristics and the complexity of these network ecosystems  \cite{maille_toward_2016} \cite{valletti_net_2016}.

Many regulatory authorities around the world assumed the task of defending fairness and universality principles for computer networks such as the Internet \cite{webb_net_2012}. These regulatory instances created rules or offered guidelines for traffic discrimination towards network neutrality \cite{berec_berec_2016} \cite{federal_communications_commission_protecting_2015} \cite{camara_dos_deputados_decreto_2016}. In general, these rules limit the ability of networks to interfere with applications, content, and services, prohibiting undue discriminatory Internet Traffic Management (ITM), also referred to as unreasonable network management for networks in general \cite{belli_discourse-principle_2013} \cite{berec_berec_2016}. NN regulatory norms and guidelines were issued and implemented at the national level by the USA\cite{federal_communications_commission_protecting_2015}, the European Community(BEREC\footnote{BEREC - \url{http://berec.europa.eu/} }) \cite{berec_berec_2016}, and Asian and Brazilian regulatory institutions\cite{camara_dos_deputados_decreto_2016} \cite{marsden_comparative_2016}.

In 2020, network neutrality gains a new push due to the pandemic situation around the world. In effect, the sudden increase of Internet usage by citizens locked in their homes and the need for communication quality and minimum bandwidth allocation assurance points to the need to regulate the Internet and have a neutral, even, and balanced resource allocation policy for networks in general.

Although the regulation defining network neutrality principles and norms have existed and the network neutrality research has been active for a decade, NN tools and technical solutions deployed in networks are under-explored.

Most research efforts and tools available so far are focused on detecting NN violations and providing methods and techniques for traffic differentiation in networks \cite{garrett_monitoring_2018}. In addition to that, NN research and discussion have strongly focused on regulatory issues, business models, and network traffic discrimination impact for users and providers \cite{bourreau_net_2015} \cite{peitz_net_2016}.

In this paper, we propose the tool NN-PCM (Network Neutrality Policy Conformance Module). NN-PCM is a network neutrality policy conformance module based on a Bandwidth Allocation Model (BAM) that accomplishes the BEREC (Body of European Regulators for Electronic Communications) network neutrality non-discriminatory policy for networks.

NN-PCM approach is new in the sense that the BAM model and configuration used to separate network traffic in classes inherently guarantees the BEREC network neutrality non-discriminatory policy for networks. In brief, the modeling approach used to group network users in classes coupled with the leveraging of the BAM model characteristics results in accomplishing the BEREC network neutrality non-discriminatory policy for networks.

Unlike the approaches in current NN tools approaches, the NN-PCM does not detect NN violations or focuses on monitoring traffic differentiation. It directly enforces the NN BEREC policy based on its operation, and that is new and differentiates this solution from the current ones available. As such, it has a new outstanding characteristic that allows NN modeling and NN deployment \emph{by design}\footnote{For the scope of this paper, NN, by design, means that the BAM characteristics inherently accomplish the network neutrality.}. To achieve such a goal, the NN-PCM leverages the characteristics of bandwidth allocation models (BAMs) and uses an SDN/OpenFlow-based architecture to program and control the network.

The organization of the paper is as follows. Section 2 presents an overview of the related works. Section 3 presents the NN-PCM motivation followed by sections 4 and 5 presenting BAM and BEREC basic concepts that serve as the primary background for the following discussion. Section 6  presents NN-PCM architecture, and section 7 develops the analytical formulation of the model and illustrates its configuration. Section 8 discusses the conformance to BEREC ITM requirements with the final considerations following in section 9.

\section{Related Work}\label{sec:relatedwork}

Network neutrality may be monitored and evaluated in the network in different ways. Garret \cite{garrett_monitoring_2018} and Li \cite{li_large-scale_2019} present an extensive survey of monitoring techniques and practices for traffic differentiation detection. These techniques can be used for the detection of traffic violations from the perspective of a user. They differ from the proposed NN-PCM solution that addresses the NN problem changing the perspective from the users to the network's side. NN-PCM addresses the NN principles and applies them to network operation and management. A NN solution based on traffic differentiation, as indicated by Garret \cite{garrett_monitoring_2018}, has enormous challenges to be deployed in networks.

Another way to treat NN in the network is to provide a technical solution or tool that, based on its operation, effectively deploys the defined NN policy.  Wójcik\cite{wojcik_net_2011} proposes a technical solution aligned with the former approach using a QoS architecture. According to the authors, the FAN (Flow-Aware Networking) assures implicit service differentiation based solely on the traffic characteristics without any possibility of undesirable interference by ISPs or Internet users. In this proposal, some control of routers is assumed by using QoS configurations that are not always easy to realize for most practical situations. The NN-PCM tool differs from this solution by deploying neutrality grouping users in classes and allocating bandwidth to them based on the BAM characteristics that, by its turn, result in a neutral operation.

Schewick \cite{van_schewick_network_2015} proposed recently eight possible network neutrality rules, but does not provide details about how these rules are modeled and implemented for networks. One of the proposed rules allows networks to treat classes of applications differently if they treat equally all traffic within each class. This specific rule is aligned with the NN-PCM model for network user grouping in traffic classes for NN conformance.

The works in Garret\cite{garrett_monitoring_2018},  Ravaiol\cite{ravaioli_towards_2015} and Zhang\cite{zhang_network_2014} detect network neutrality violations through end-to-end measurements using various techniques like censorship, traffic differentiation and traffic modification. These approaches detect NN violations, but do not effectively deploy the NN police. This is fundamentally different from NN-PCM tool that implements the NN policy \emph{by design} controlling the allocation of bandwidth for network users.

The work in Maltinsky\cite{maltinsky_network_2017} explores the security vulnerabilities of the network neutrality measurements and violations detection techniques as far as they expose certain types of traffic. The NN-PCM NN deployment by design offers an advantage in this respect as far as the traffic is not exposed to any tool.

Mayer\cite{mayer_framework_2018} discusses the legitimacy of monitoring methods used by ISPs and suggests an open-source framework that explores a certain number of measurements and points out the need for transparency in how ISPs monitor and discriminate user traffic. The NN-PCM provides legitimacy and  transparency by exposing and advertising the user's traffic class mapping used in the network. Therefore, the NN-PCM provides configuration visibility for monitoring and auditing purposes for all stakeholders and players involved. This is a fundamental NN-PCM advantage in terms of facilitating NN regulation.

The work in Sivaraman\cite{sivaraman_opentd_2019} is similar to NN-PCM.  It is based on classes of traffic, and for each class, there is an allocated bandwidth. NN-PCM fundamentally differs from this work by allocating bandwidth dynamically on a per-demand basis, resulting from the inherent behavior of the bandwidth allocation model (BAM) used. In summary, the \emph{by design} characteristics of NN-PCM allow a more flexible and dynamic configuration and operation for the network.

Most recently developed works on NN discuss market implications, social impacts, and regulatory issues like innovation, competition, acceptable forms of network management, price and service differentiation, and the importance of traffic management \cite{stocker_state_2020} \cite{bauer_complementary_2018} \cite{glass_new_2019} \cite{abiteboul_transparency_2019} \cite{ma_routing_2017} \cite{bourreau_net_2015} \cite{belli_discourse-principle_2013} \cite{choi_net_2015}. Although highly relevant from the societal point of view, these works also do not discuss or propose any technical solution or tool to handle the network traffic and to provide neutrality.

To the best of our knowledge, there is not yet a proposed solution that models user's traffic and accomplishes the BEREC network neutrality policy in a transparent, non-discriminatory, exceptional, and proportional way.

\section{Motivation}\label{sec:motivation}

The primary motivation of the NN-PCM is to fill an existing gap and lack of tools for deploying the BEREC network neutrality policy in networks.

NN-PCM models and enforces BEREC NN. It does not focus on detecting violations or measuring traffic characteristics, and, in effect, it deploys the NN policy while granting bandwidth for network users. These characteristics go beyond the currently available type of tools that have been extensively explored in the domain, and that simply detect NN violations without providing a solution, as to enforce network neutrality rules to accomplish network neutrality.

NN-PCM can be used by corporate networks, ISPs, and Content Providers. It is a tool that groups network users in traffic classes and deploys the network neutrality business model based on the inherent behavior of the bandwidth allocation model (BAM).

Bandwidth allocation models are used to implement the NN-PCM. Therefore, two main questions must be discussed initially:
\begin{itemize}
    \item How bandwidth allocation models work; and
    \item How BAM models characteristics are used to implement the NN-PCM and to achieve NN.
\end{itemize}

These issues are considered in the following sections.

\section{Bandwidth Allocation Models - Summary} \label{sec:BAMOperation}

\begin{figure} [ht]
\centering
\includegraphics[width=0.5\textwidth]{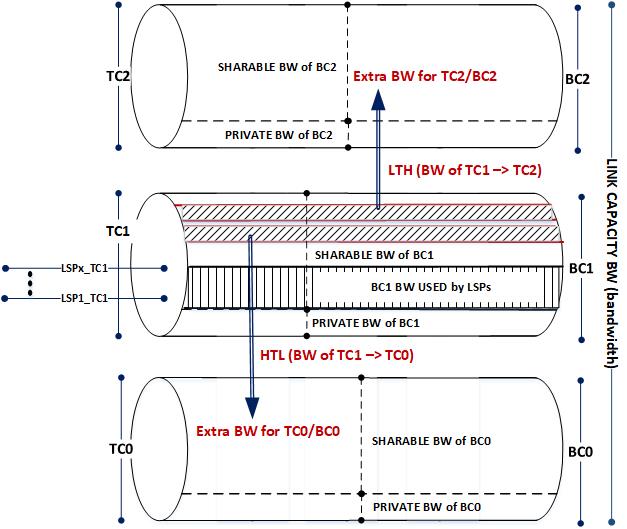} 
\caption{BAM generalized bandwidth allocation and sharing methods\cite{reale_orquestracao_2019}}
\label{fig:BAMOperation}
\end{figure}

Bandwidth allocation models define a resource allocation and resource sharing method.  Reale\cite{reale_g-bam:_2014} \cite{reale_orquestracao_2019} discusses BAM operation, and Figure \ref{fig:BAMOperation} summarizes all the allocation sharing methods that are possible. With BAM, bandwidth can be private or can be shared between high and low priority users in both directions (LTH - Low to High and HTL - High to Low sharing), depending on the BAM model used.

The most basic BAM models (MAM - Maximum Allocation Model\cite{faucher_maximum_2005} and RDM - Russian Dolls Model\cite{le_faucheur_russian_2005}) are resource sharing theoretical models described by RFCs (Request for Comments) that define how to share and to enforce different bandwidth constraints for different classes of traffic aiming traffic engineering. More recently, hybrids of the basic models were proposed \cite{tata_cam:_2013} \cite{da_costa_pinto_neto_adapt-rdm_2008}. The ATCS model (AllocTC-Sharing)\cite{reale_alloctc-sharing:_2011} is a new basic BAM model that allows resource sharing among all traffic classes and the GBAM model\cite{reale_g-bam:_2014} (Generalized BAM) generalizes the deployment of all possible resource sharing strategies for all existing BAM models.

BAMs are deployed nowadays in domains like link bandwidth sharing\cite{bahnasse_smart_2018}\cite{oliveira_cognitive_2018}, last mile networks\cite{sadon_dynamic_2012}, 5G networks with virtual network embedding\cite{trivisonno_network_2015} and fiber optic networks\cite{duraes_evaluating_2017}\cite{hesselbach_management_2016}.

 BAM is mostly applied to networks in which resources such as link bandwidth and fiber optic slots (lambda circuits) are limited. In this paper, BAMs are used to allocated bandwidth \footnote{The term \textbf{bandwidth}, for the scope of this paper, will mean the \textbf{capacity} or \textbf{transmission rate} allocated to a user in a link.} to users in network links.

BAM operation requires 3 main configuration steps (Figure \ref{fig:BAMOperation}) \cite{reale_g-bam:_2014} \cite{reale_orquestracao_2019}:
\begin{itemize}
    \item The grouping of applications into traffic classes (TCs) with similar network requirements;
    \item The configuration of the maximum amount of bandwidth (resource) per traffic class (Bandwidth Constraints - BC); and
    \item The definition of the strategy for bandwidth (resource) sharing among traffic classes (TCs) such as private, LTH (Low-To-High), and HTL (High-To-Low) sharing \cite{reale_orquestracao_2019}.
\end{itemize}

There are 3 basic BAM models, described in Reale\cite{reale_g-bam:_2014} \cite{reale_alloctc-sharing:_2011} and  Neto\cite{da_costa_pinto_neto_adapt-rdm_2008}:
\begin{itemize}
    \item Maximum Allocation Model (MAM)\cite{faucher_maximum_2005};
    \item Russian Dolls Model (RDM)\cite{le_faucheur_russian_2005}; and
    \item AllocTC-Sharing (ATCS)\cite{reale_alloctc-sharing:_2011}.
\end{itemize}

Maximum Allocation Model (MAM) allocates bandwidth without any bandwidth sharing between traffic classes (TCs). In summary, MAM traffic classes have exclusive use of their bandwidth resources (BCs).

In the Russian Dolls Model (RDM), it is possible to share bandwidth not used by high priority classes. As such, low priority classes may have access to extra bandwidth in addition to their bandwidth resources (Figure \ref{fig:BAMOperation}) (HTL - High to Low sharing).

Finally, the AllocTC-Sharing model (ATCS) generalizes bandwidth sharing among all classes (low and high priority). In this case, all unused bandwidth might be shared among classes independently of their priority (Figure \ref{fig:BAMOperation}) (HTL - High to Low and LTH - Low to High sharing).

The ATCS model is adopted in the NN-PCM since it includes and reproduces the operation of all BAM basic models (MAM and RDM) and supports a broader scope of applications\cite{reale_analysis_2014}.

\section{The BEREC Reasonable Internet Traffic Management}\label{subsec:ITM}

BEREC (Body of European Regulators for Electronic Communication) is the European body that brings together all national regulatory authorities (NRAs) and explores such issues as transparency, competition, quality of service, quality monitoring, and IP interconnection in the context of network neutrality \cite{body_of_european_regulators_for_electronic_communications_all_2020}.

As discussed in BEREC\cite{berec_berec_2016}, the BEREC reasonable Internet Traffic Management (ITM) is an acceptable practice that can be used by networks to implement non-discriminatory rules and policies. The so-called reasonable-ITM-exception is discussed in regulatory documents \cite{berec_berec_2016} \cite{federal_communications_commission_protecting_2015} \cite{camara_dos_deputados_decreto_2016} and studies \cite{stocker_state_2020}  \cite{belli_net_2016} \cite{van_schewick_network_2015}. In summary, it is a fair and justifiable deviation from the strict NN non-discrimination rule, where all traffic is absolutely equal in terms of network requirements \cite{belli_value_2013}  \cite{belli_discourse-principle_2013}.

The reasonable ITM is a concept proposed by Europeans through BEREC \cite{belli_net_2016}. There is a long-term discussion on what network neutrality concept should be used for all actors involved that we consider as out of the scope of this paper. Indeed, we consider BEREC ITM proposal a valid reference for network neutrality based on two main reasons:
\begin{itemize}
   
    \item The agnostic approach inherently embedded in the BEREC ITM does not significantly differ from concepts supported by other regulatory institutions; and
     \item BEREC proposal is based on objective technical requirements \cite{belli_net_2016}.
\end{itemize}

In terms of network neutrality deployment, ITM adopts technical practices and management approaches whose purpose is to maintain, protect, and ensure an efficient network operation \cite{van_schewick_network_2015}. Efficient operation means that the NN solution must take into account that different applications might have different and conflicting network requirements. NN ITM exception considers it acceptable to adopt management practices that deal with different application requirements.

Necessarily, to qualify a management practice as conforming to ITM, it must meet the following requirements \cite{berec_berec_2016} \cite{barreto_neutralidade_2017}:

\begin{enumerate}
\item Have a legitimate network management purpose;\label{req1}
\item Be transparent;\label{req2}
\item Be non-discriminatory;\label{req3}
\item Be proportional;\label{req4} and
\item Be exceptional.\label{req5}
\end{enumerate}

The NN-PCM prototype architecture, configuration, and operation that lead to the conformance with these ITM requirements are presented and evaluated next.

\section{The NN-PCM Architecture Model}

The NN-PCM (Network Neutrality Policy Conformance Module) objective is dual fold:

\begin{itemize}
    \item To manage the allocation to users of link bandwidth in the network using a bandwidth allocation model (BAM); and
    \item To accomplish the BEREC Reasonable Internet Traffic Management (ITM) concomitantly.
\end{itemize}

\begin{figure} [ht]
\centering
\includegraphics[width=0.5\textwidth]{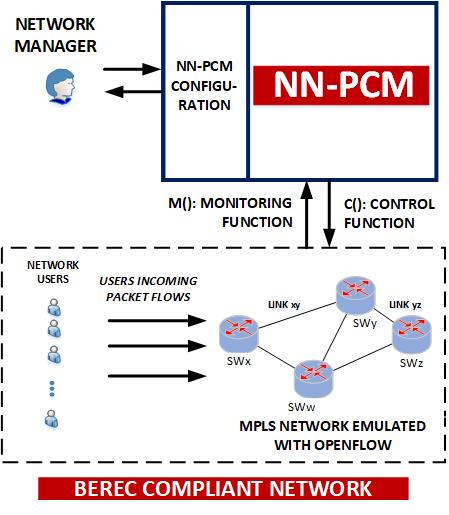} 
\caption{NN-PCM architecture main components and interfaces}
\label{fig:NNPCMInterfaces}
\end{figure}

The main components and interfaces of the NN-PCM prototype architecture is illustrated in Figure \ref{fig:NNPCMInterfaces}. They are the NN-PCM module, the controlled network, users generating traffic flows, and a set of interfaces for the network manager and monitoring and controlling the target network. The NN-PCM adopts a centralized network control operation using the OpenFlow protocol under the Software-defined Network (SDN) paradigm \cite{kreutz_software-defined_2014}.

The NN-PCM interacts with the controlled network using two basic functionalities:
\begin{itemize}
    \item  $M()$ for monitoring the network concerning new incoming flows; and
    \item $C()$ to control the network by programming flows deployment over the set of switches (Figure \ref{fig:NNPCMInterfaces}).
\end{itemize}

The NN-PCM configuration (Figure \ref{fig:NNPCMInterfaces}) is another basic element of the NN-PCM architecture. It is the configuration interface that allows the definition of the adopted BAM, its configuration and the grouping of network users in classes of traffic according to their network requirements.

\begin{figure} [ht]
\centering
\includegraphics[width=0.5\textwidth]{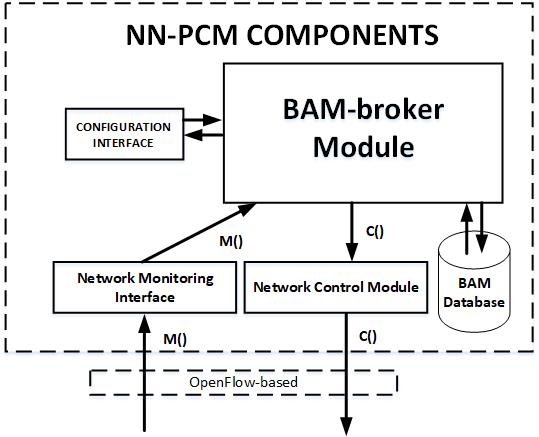} 
\caption{The NN-PCM module main blocks and components}
\label{fig:NNPCIArchitecture}
\end{figure}

The main blocks and components of the NN-PCM module are illustrated in Figure \ref{fig:NNPCIArchitecture}. It is composed by:
\begin{itemize}
    \item The BAM-broker module;
    \item The monitoring interface;
    \item The network control module (NCM); and 
    \item An interface to allow NN-PCM configuration.
\end{itemize}

The BAM module acts as a bandwidth broker and assures the fair and neutral network operation while allocating bandwidth to flows generated by the users. Even when the input traffic fluctuates and bandwidth resources are exhausted, the BAM dynamic behavior allows a neutral distribution of bandwidth under the BEREC policy.

The network monitoring interface (NMI) is responsible for catching and forwarding to the BAM-broker the new bandwidth requests for the new incoming flows in the network. The new input flow is detected by the OpenFlow switches when a packet arrives in the switch, and there is no defined entry for that flow in its flow-table. When this happens, a \textit{PacketIn} OpenFlow message is sent to the NN-PCM that acts as an SDN/OpenFlow network controller.

The network control module (NCM) is the interface through with OpenFlow commands are sent to the network to configure, program, and manage flows in the switch's flow-tables.

The BAM database stores all information concerning the currently configured flows in the network for all users, switches, and paths. It is used by BAM operation to decide on the allocation of bandwidth to new incoming flows in the network.

\begin{figure} [ht]
\centering
\includegraphics[width=0.5\textwidth]{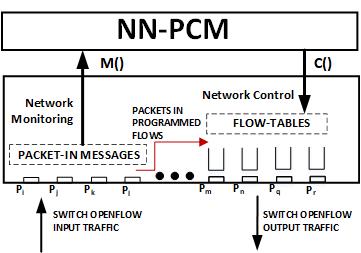} 
\caption{OpenFlow switch monitoring and control}
\label{fig:SwitchOperation}
\end{figure}

The internal operation of the OpenFlow switch in the network is monitored and controlled, as illustrated in Figure \ref{fig:SwitchOperation}. New flows not yet configured in the switch flow-table generate a \textit{PacketIn} message that is sent to the NN-PCM. A new end-to-end flow (source to destination) is then created for the network by configuring a new flow entry at the flow-table for all the switches in the flow path.

The OpenFlow-based centralized monitoring and control of the network allow the NN-PCM to inherit the advantages of an SDN-based deployment. The centralized monitoring and control include a single view of the entire network and the capability to dynamically configure any switch in the network according to the current traffic profile.

\begin{figure} [ht]
\centering
\includegraphics[width=0.5\textwidth]{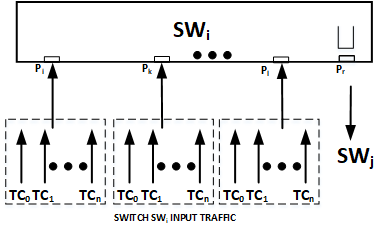} 
\caption{BAM operation by switch and by link}
\label{fig:BAMAnalytical}
\end{figure}

The NN-PCM architecture also solves another essential problem present in network neutrality management deployments that is where to place the management control. In NN-PCM, the NN management is effectively positioned at the incoming traffic switch for all switches in the network (Figure \ref{fig:BAMAnalytical}) (Equation \ref{eq:LSP}). This is achieved by the basic OpenFlow monitoring and programming operation. NN-PCM requires new flow identification, done by the OpenFlow input traffic switches. OpenFlow switch programming includes LSP flows being programmed in terms of bandwidth and queues, and NN-PCM does this for all switches in the LSP path \cite{mendiola_survey_2017}.

The NN-PCM architecture is the base for the analytical NN-PCM description and the BAM behavior simulation analysis discussed in the following sections.

The next question concerning the NN-PCM is how applying the BAM operation to a set of flows and traffic classes in a network conforms (or not) to the BEREC network neutrality policy. This issue is addressed in the following section.

\section{Accomplishing Conformance to BEREC ITM Policy with the BAM}

The fundamental conception premise of the NN-PCM is that the BAM operation does accomplish conformance to the NN BEREC ITM requirements defined in BEREC\cite{berec_berec_2016}. Concerning the ITM requirements, this means that:

\begin{enumerate}
\item The BAM operation has a \textbf{legitimate network management purpose};\label{req1}
\item Its configuration and operation is \textbf{transparent};\label{req2}
\item It has a \textbf{non-discriminatory} behavior concerning network users;\label{req3}
\item It has a \textbf{proportional} behavior;\label{req4} and
\item It adopts an \textbf{exceptional} behavior when it is acceptably required.\label{req5}
\end{enumerate}

This ITM compliant behavior is evaluated in the next sections but, firstly, we introduce the basic design neutrality principle adopted by the NN-PCM, present an analytical formulation of the BAM operation and illustrate the configuration of NN-PCM operation.

\subsection{NN-PCM Neutrality Model}

The general principles used to model a neutral network operation with the NN-PCM are the following:

\begin{itemize}
    \item The BAM model operation accomplishes a per-link neutral bandwidth allocation; and
    \item Users with equivalent network requirements are grouped into the same traffic classes (TCs) and are subject to non-discriminatory management.
\end{itemize}

In addition to the above principles, the following definitions hold for the NN-PCM operation:

\begin{itemize}
    \item The BAM model manages each link (attached to a switch port) in the physical network independently (Figure \ref{fig:BAMAnalytical});
    \item A network path (LSP from source to destination) is composed of various links. The BAM manages the network path by managing the set of links belonging to the path independently (Equation \ref{eq:LSP}); and
     \item Accomplishing link neutrality implies accomplishing path neutrality by having the BAM model managing all links belonging to the path independently.
\end{itemize}

\subsection{NN-PCM BAM Analytical Formulation}

The BAM solves effectively a traffic engineering problem and, in the NN-PCM implementation, it is as a centralized controller operating according the SDN paradigm with the OpenFlow protocol.

The network infrastructure is modeled as a bidirectional physical graph ${\aleph} ^q= \left( SW^q, L^q\right)$, where $SW^q$ = $\{sw_1, sw_2, sw_3, ..., sw_n\}$ is the set of OpenFlow switches in the target network $q$ and $L^q$ = $\{l_{ij}\}$, with $i$ and $j$ = $1, 2,3, ..., n$, is the set of physical links, with link $l_{ij}$ connecting $sw_i$ to $sw_j$.

The switch connectivity matrix (SCM) is defined by $C=[c_{ij}]$ with $i$ and $j$ $=1,2,3,..., n$, and  $c_{ij}=1$ if $sw_i$ is connected to $sw_j$; $0$ otherwise and $c_{ii}=0$ $\forall i$ by definition.

The link bandwidth matrix (LBM) for the physical network is defined by $LB=[LB_{ij}]$ with $i$ and $j$ $=1,2,3,..., n$, where $LB_{ij}$ is the total bandwidth of the link $l_{ij}$ managed by the BAM to allocate all incoming traffic arriving at $sw_i$ to $sw_j$.

The BAM model defines for each link $l_{ij}$ a set of traffic classes $(TC^k)$, with $k$ $=0,1,..., n$. Each traffic classes $TC^k_{ij}$ belonging to the link $l_{ij}$ has a bandwidth $BC^k$ (Bandwidth Constraint), such that:
\begin{equation}
LB_{ij}\ge\sum_{i=1}^{n}\sum_{j=1}^{n}\sum_{k=1}^{n} BC^k_{ij}
\end{equation}

In the BAM operation by switch and by link, $T_{ij}TC^kP_{sw_i}^z$ is the incoming traffic at port $z$, with $z=1,2,..,n$, of $sw_i$, belonging to traffic class $TC^k$, with $k=0,1,...,n$, whose destination is $sw_j$ through link $l_{ij}$ (Figure \ref{fig:BAMAnalytical}).

The traffic class (TC) used bandwidth (TCUB) matrix is defined by $TCUB$$=[TC^{k}{UB_{ij}}]$, with $k=0,1,...,n$, and $i$ and $j$ $=1,2,..,n$.

The bandwidth constraint for the traffic classes is:
\begin{equation}
TC^kUB_{ij}\le BC^k_{ij} \forall k,i,j
\end{equation}

A LSP is defined as set of $Z$ 2-switch directed graph ${\aleph}^Z$ $= ( SW^Z, L^Z)$, $Z=1,2,...,z$, where $SW^Z=\{sw^z_s,sw^z_d\}$ are the source and destination switches on the $z^{th}$ LSP section  and $L^Z=\{l^z_{sd}\}$ is the link connecting source to destination switches on the $z^{th}$ LSP section with bandwidth allocated $BA_{U_x}TC^k_{sd}LSP_m$ for each link.

For the sake of simplicity to reduce simulation complexity, the LSP path computation is made available to NN-PCM by the LSP connectivity path ($LSP_{CP}$) matrix:
\begin{equation} \label{eq:LSP}
\begin{split}
LSP_{CP}=
[\sum_{\alpha=1}^{n}\overbrace{(LSP_{{seg}_{mn}}\parallel\cdots\parallel(LSP_{{seg}_{yz}})}^{\alpha}]
\end{split}
\end{equation}

The LSP has source switch $sw_m$ and destination switch $sw_z$ and is composed by concatenated segments with interconnected switches and links on the path as follows:
\begin{equation}
LSP_{{seg}_{xy}}=(sw_x,sw_y,l_{xy})
\end{equation}

Computing the LSP path for multiple incoming traffic is a complex traffic engineering optimization problem that has no impact on the evaluation of the network neutrality conformance.

The bandwidth allocated $BA_{U_x}TC^k_{sd}LSP_m$ for the LSP segments $LSP_{{seg}_{xy}}=(sw_x,sw_y,l_{xy})$  has a fixed value for a given traffic class. This approach is consistent with the principle that users with similar bandwidth requirements are mapped to the same traffic class and have a non-discriminatory treatment.

\subsection{ATCS Model Analytical Formulation}

The AllocTC-Sharing (ATCS) model \cite{}  used in the NN-PCM has additional capabilities concerning a basic BAM model:

\begin{itemize}
    \item Traffic classes (TCs) have a priority; and
    \item Bandwidth sharing is allowed between all traffic classes, and the amount of bandwidth shared is limited.
\end{itemize}

The ATCS traffic class priority for all links $l_{ij}$ with $i,j = 1,2, ..., n$ is:
\begin{equation} \label{eq:pri}
TC_p^0 > TC_p^1 > TC_p^2 > \dots > TC_p^n
\end{equation}

In the BAM convention, traffic class $0$ priority ($TC_p^0$) has the highest value, and traffic class $n$ priority ($TC_p^n$) the lowest value.

The traffic class priority, as described in Reale\cite{reale_alloctc-sharing:_2011}, is used by the ATCS to:

\begin{itemize}
     \item Allocate unused bandwidth from other traffic classes; and
     \item Preempt or return shared bandwidth previously allocated from other traffic classes.
  
\end{itemize}

ATCS policy allows each traffic class $TC^k$ to have two bandwidth partitions: private and public. The private bandwidth is exclusively allocated for $TC^k$ users, and the public bandwidth may be allocated by other traffic classes when not used by  $TC^k$ users. As such for all $l_{ij}$ with $i,j = 1,2, ..., n$:
\begin{equation}
TC^kB_{ij}= TC^k_{pri}B_{ij}+TC^k_{pub}B_{ij}
\end{equation}

The traffic class used bandwidth (TCUB) matrix for ATCS must then be rewritten as follows for all $l_{ij}$ with $i,j = 1,2, ..., n$:
\begin{equation} \label{eq:TCUB}
\begin{split}
TC^kUB_{ij}= [TC^k_{pri}UB_{ij}+TC^k_{pub}UB_{ij} \\
+\sum_{z=1}^{n}TC^k_{pub}UB_{ij}]
\forall z\neq k
\end{split}
\end{equation}

Equation \ref{eq:TCUB} describes the ATCS behavior. ATCS model can allocate unused public bandwidth from other traffic classes. ATCS loan bandwidth from other traffic classes considering the available public bandwidth from the lower to the highest priority TCs.

\begin{figure} [ht]
\centering
\includegraphics[width=0.45\textwidth]{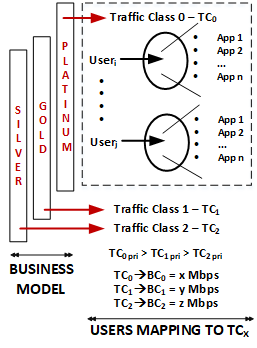} 
\caption{NN-PCM network neutrality model and BAM configuration}
\label{fig:NNPCM-TC}
\end{figure}

Users are mapped to traffic classes $\sum_{k=1}^{n}TC^k$ (Figure \ref{fig:NNPCM-TC}). As such, for all links $l_{ij}$ with $i,j = 1,2, ..., n$, $TC^k$ has a set of users $\sum_{1=1}^{n}U^i$.

In the NN-PCM current evaluation, each $U^i$ may have any number of LSPs over the network. The limit of LSP setup is only limited by the amount of private and public bandwidth by class and by link.

In summary, for each traffic class $TC^k$, there is a User LSP matrix (ULSP) with LSPs (Equation \ref{eq:LSP}) conveying user application traffic from source switch $sw_s$ to destination destination $sw_d$ (Figure \ref{fig:NNPCM-TC}):

\begin{equation} \label{eq:UTOTC}
\begin{split}
ULSP=[\sum_{y=1}^{m}U^y\sum_{x=1}^{n}LSP^x]\forall TC^k, k=0,1,\dots,n.
\end{split}
\end{equation}

\subsection{NN-PCM Configuration}

The NN-PCM is configured by mapping network users, services, and applications, from now on, generically referred as \textbf{users}, into a set of BAM traffic classes (TCs) (Figure \ref{fig:NNPCM-TC}).

There are various alternatives to map users to traffic class and this mapping strongly depends on the business model adopted by the network owner. Just as an illustration, Figure \ref{fig:NNPCM-TC} shows 3 TCs (platinum, gold, and silver) where platinum has the highest and silver has the lowest priority (Equation \ref{eq:pri}).

From the network neutrality perspective, to group users into classes allows:
\begin{itemize}
    \item To have users with equivalent bandwidth requirements in the same class; and 
    \item To have users with either distinct bandwidth requirements or distinct priorities in different classes. 
\end{itemize}

From the network manager perspective, the mapping of users into traffic classes (TCs), has to consider two essential aspects:
\begin{itemize}
    \item The user bandwidth requirement for the application or service execution; and
    \item The network policy ruling its business, managerial practices, and legal constraints.
\end{itemize}

 As another example to illustrate users mapping alternatives, certain users need guaranteed minimum bandwidth, while others do not. Besides, some users  take advantage of additional bandwidth, if available, to improve its user experience. These criteria are used by the network manager to map users with similar requirements in the same class.

It is relevant to indicate that the mapping of users on a traffic class (TC) does not guarantee bandwidth for all TC users. Network resources are, in most real cases, scarce and disputed. The NN-PCM supports bandwidth disputes under congestion, with the BAM model assuring that the NN principle is preserved.

The NN-PCM configuration can be summarized as follows:
\begin{itemize}
    \item Map users with similar network requirements on a set of traffic classes (TCs);
    \item Configure the available bandwidth per class (bandwidth constraints) for each TCs; and
    \item Choose the BAM model (MAM, RDM, ATCS, other) that defines the behavior for bandwidth sharing among TCs.
\end{itemize}

In practice, there are various possibilities to group users into TCs and configure the TCs bandwidth (BCs). The evaluation of these alternatives is out of the scope of this paper.

Table \ref{tableI} illustrates one of many possible users grouping into classes. This grouping is based on the association of QoS classes as proposed in ITU-T Recommendation Y.1541 and the transfer capabilities requirements extracted from ITU-T Recommendation Y.1221 \cite{itu-t_y.1541:_2002} \cite{itu-t_y.1221:_2002}.

\begin{table*}[h!]
 \centering
 {\renewcommand\arraystretch{1.2}
 \setlength{\arrayrulewidth}{0.1mm}
 \caption{Example of Users to Traffic Class (TC) Mapping}
 \label{tableI}
 \begin{tabular}{ l l l }
  \cline{1-1}\cline{2-2}\cline{3-3}  
    \multicolumn{1}{|p{7cm}|}{ USERS \centering} & \multicolumn{1}{p{1.5cm}|}{ TC \centering } & \multicolumn{1}{p{5cm}|}{ TRAFFIC \centering }\\ 
    
  \cline{1-1}\cline{2-2}\cline{3-3}  
    \multicolumn{1}{|p{7cm}|}{ \text{Medical images, virtual reality, emergency, ...} \centering } & \multicolumn{1}{p{1.5cm}|}{ \text{TC0} \centering } & \multicolumn{1}{p{5cm}|}{ \text{Critical and emergency} \centering }\\ 
    
  \cline{1-1}\cline{2-2}\cline{3-3}  
    \multicolumn{1}{|p{7cm}|}{ \text{Audio/video, real-time applications, ...} \centering } & \multicolumn{1}{p{1.5cm}|}{ \text{TC1} \centering } & \multicolumn{1}{p{5cm}|}{ \text{Real time; bandwidth intensive} \centering }\\
    
  \cline{1-1}\cline{2-2}\cline{3-3}  
    \multicolumn{1}{|p{7cm}|}{ \text{Web applications, browsing, file transfer, ...}\centering } & \multicolumn{1}{p{1.5cm}|}{ \text{TC2} \centering } & \multicolumn{1}{p{5cm}|}{\text{Best effort; non real-time} \centering }\\  
  \hline

 \end{tabular} }
\end{table*}

\section{BAM Conformance to ITM Requirements - Discussion and Proof-of-Concept}

In this section, we demonstrate how the NN-PCM operating with a BAM model does conform to the 5 previously defined criteria for BEREC reasonable Internet Traffic Management (ITM).

As presented in Reale\cite{reale_g-bam:_2014}, the BAM models have two types of behavior \footnote{The BAM behavior is the set of characteristics that the BAM model has when it allocates bandwidth for users}:
\begin{itemize}
    \item BAM general behavior; and
    \item BAM specific behavior.
\end{itemize}

The general behavior corresponds to the set of BAM operational characteristics that are valid for all BAM models. The specific BAM behavior, as the name suggests, corresponds to the individual operational characteristics of each BAM model. As an example, MAM and RDM have specific behaviors, since MAM does not share resources while RDM does.

In summary, the general BAM behavior certifies conformance to ITM requirements 1, 2, 3, and the specific BAM behaviors certify conformance to ITM requirements 4 and 5.

The conformance to ITM requirements 1, 2, 3 is based on the interpretation of the ITM requirements in relation to the general behavior of BAM models. The conformance to ITM requirements 4 and 5 depends on the evaluation of BAM specific behavior by simulation. As such, we first present the simulation scenario used, followed by the conformance analysis to all ITM requirements.

An inference generalization holds for the NN-PCM ITM conformance evaluation. In effect, it is necessary to evaluate the ITM conformance just for one network link. That is so because the BAM model manages bandwidth allocation for all links of the network independently. In a network with N links, N independent BAM instances manage the available resources, one for each link. This means that the same general and specific behaviors exist for all links in the network if the same BAM model is configured for all links.

Based then on the premises that the BAM control links independently and the same BAM model is used for all links of the network, the simulation, and verification of the NN-PCM ITM conformance for any link of the network, leads to the conclusion that it conforms for all links and conform for any path of the network. Consequently, the NN-PCM is ITM conform for the entire network.

\subsection{Simulation Network for BAM Behavior Analysis}

\begin{figure} [ht]
\centering
\includegraphics[width=0.45\textwidth]{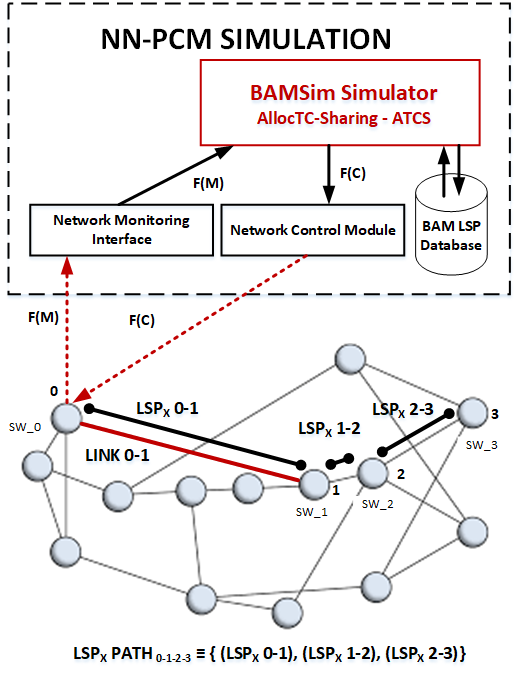} 
\caption{NN-PCM simulation set-up}
\label{fig:NNPCM-Simulation}
\end{figure}

The simulated network is an MPLS (MultiProtocol Label Switching) network in which users request circuits (LSP - Label Switched Path) passing through network links (Figure \ref{fig:NNPCM-Simulation}). The requested resource in all links over the LSP path is the bandwidth required by the requested LSP. The requested LSP belongs to a user, configured by the manager in a traffic class (TC0, TC1 or TC2).

From the user (external) point of view, the NN-PCM acts as a broker that emulates an MPLS network. That means, in summary, that users get an LSP that is fully emulated by the NN-PCM through its SDN/OpenFlow-based control functions ($M()$ and $C()$) (Figure \ref{fig:NNPCIArchitecture}) with no need to use MPLS signaling protocols.

The basic simulation definitions are:
\begin{itemize}
     \item The simulator used is the BAMSim (BAM Simulator) \cite{reale_g-bam:_2014}; and
     \item NN-PCM uses the AllocTC-Sharing (ATCS) BAM model \cite{reale_alloctc-sharing:_2011}.
   
\end{itemize}

BAMSim has features to simulate various BAM behaviors using the NSF (National Science Foundation) benchmark topology with 14 nodes and 42 bidirectional links. For the scope of NN-PCM specific ITM conformance evaluation, the link 0-1 is considered (Figure \ref{fig:NNPCM-Simulation}).

The simulation runs use two network traffic scenarios.
\begin{itemize}
    \item The first scenario uses random input traffic requesting LSPs and evaluates whether the NN-PCM achieves a non-discriminatory resource distribution in a proportional way (ITM requirement 4).
     \item The second scenario uses a previously established (non-random) input traffic profile to test whether the NN-PCM operation is exceptional (ITM requirement 5).
\end{itemize}

The simulation configuration parameters are:
\begin{itemize}
    \item The network link simulated (Figure \ref{fig:NNPCM-Simulation}) has a total bandwidth of 1000 Mbps between nodes 0 and 1 of the topology;
    \item Sharing limit: 100\% of BC; and
    \item 3 Traffic Classes: TC0 with BC0 = 25\%, TC1 with BC1 = 35\% and TC2 with BC2 = 40\% of the total link bandwidth.
\end{itemize}

The simulation (scenarios 1 and 2) runs for 5 hours with 5 phases of 1 hour. Other simulation parameters are:
\begin{itemize}
    \item LSP bandwidth randomly distributed between 5 and 15 Mbps;
    \item Interval of LSP arrival requests modeled exponentially;
    \item Exponentially modeled LSP time life - average of 300 seconds; and
    \item Halting criteria - 5 hours.
\end{itemize}

\subsection{ITM Requirement 1 Conformance Accomplishment - Legitimate Management Purpose}

ITM requirement 1 defines that the network management practice must have a \textbf{legitimate network management purpose}. By legitimate network management purpose, it is meant that the applied management has to be designed to maintain, protect, and ensure an efficient network operation.

The NN-PCM, being based on any BAM, is conform to ITM requirement 1 due to the BAM's general behavior. The BAM's general behavior has a legitimate network management purpose because it maintains and protect bandwidth for traffic classes (TCs) and, also, allows efficient use of the available bandwidth, responding to different technical requirements of specific categories of traffic configured in different traffic classes (TCs) \cite{reale_analysis_2014}.

In summary, each TC has its bandwidth maintained, protected from other external users, and, in exceptional circumstances, it is allowed bandwidth sharing leading to efficient network operation \cite{reale_analysis_2014}.

\subsection{ITM Requirement 2 Conformance Accomplishment - Transparency}

The ITM requirement 2 defines that traffic management measures should be transparently included in the contract and published with a clear and comprehensive explanation of traffic management measures applied.

The ITM \textbf{transparency} requirement is accomplished by publishing or making public the NN-PCM configuration parameters. This, in practice, corresponds to the network neutrality policy implementation being defined for the target network by the manager. Various NN configuration parameters that can be made public and transparent, such as:
\begin{itemize}
    \item The number of traffic classes (TCs), the mapping of users, the bandwidth available per traffic class, and the limit for bandwidth sharing.
\end{itemize}

This information can be made transparent for users,  stakeholders, partners, and regulators.

\subsection{ITM Requirement 3 Conformance Accomplishment - Non-discriminatory} \label{sec:non-discriminatory}

ITM requirement 3 defines that the network management practice adopted has to be non-discriminatory concerning users. As discussed in Belli\cite{belli_net_2016} and Schewick\cite{van_schewick_network_2015}, non-discriminatory means basically that the management practice is application-agnostic. As stated in Schewick\cite{van_schewick_network_2015}, \textquotedblleft the management practice may have information about the application but does not make any distinction among data packets based on this information \textquotedblright. Likewise, the ITM concept establishes that traffic and users may be grouped into categories that should reflect their technical requirements \cite{berec_berec_2016} \cite{belli_net_2016}.

In the NN-PCM, users with similar network requirements  receive similar treatment. In other words, users with similar requirements are handled agnostically in the same traffic class (application-agnostic operation). Likewise, users with distinct network requirements are handled by distinct traffic classes (TCs).

As such, the operation of the NN-PCM for all possible BAM models is non-discriminatory. It is non-discriminatory because it always grants (or not) bandwidth for a group of users organized into classes with common or similar requirements. This assertive is discussed and reinforced again in the following sections when simulating the operation of the NN-PCM with a specific BAM model (AllocTC-Sharing - ATCS \cite{reale_alloctc-sharing:_2011}).

\subsection{ITM Requirement 4 Conformance Accomplishment - Proportionality}

The ITM proportionality requirement defines that we should have pieces of evidence that the network management practice reaches its defined purpose, and the actions must be necessary to reach the purpose, interfering as less as possible \cite{berec_berec_2016}.

As a practical example in the context of the NN-PCM, ITM proportionality requires that blocking, preemption, and devolution events must be proportional, that is, should not occur more than necessary to reach the purpose of allocating bandwidth to users according to their requirements as well to optimize network performance.

An ITM management practice is considered proportional when there is no less interfering and equally effective alternative to manage user's demand with the available network resources \cite{berec_berec_2016}.

In the FRFS bandwidth allocation management practice, users share the link bandwidth without any class grouping or priorities with no interference at all among them. FRFS is the best possible approach in terms of non-interfering criteria.

Based on the above ITM proportionality definition, we evaluated the NN-PCM proportionality conformance comparing the NN-PCM with the FRFS method as follows:

\begin{itemize}
    \item NN-PCM allocates bandwidth using AllocTC-Sharing BAM (ATCS) using simulation scenario 1; and
    \item The First-Requested First-Served (FRFS) allocates bandwidth using the same simulation scenario 1 for comparison.
\end{itemize}

The ITM proportionality requirement is evaluated using two network performance parameters: i) link utilization; and ii) blocking rate. The simulation results for these parameters are presented in Table \ref{tableII}.

\begin{table*}[ht]
 \centering
 {\renewcommand\arraystretch{1.2}
 \setlength{\arrayrulewidth}{0.1mm}
 \caption{ITM Proportionality Simulation Results Summary  - NN-PCM vs FRFS}
 \label{tableII}
 \begin{tabular}{ l l l }
  \cline{1-1}\cline{2-2}\cline{3-3}  
    \multicolumn{1}{|p{4cm}|}{\textbf{ PARAMETER} \centering} &
    \multicolumn{1}{p{2.5cm}|}{\textbf{ NN-PCM} \centering } &
    \multicolumn{1}{p{2.5cm}|}{\textbf{ FRFS} \centering }
  \\  
  \cline{1-1}\cline{2-2}\cline{3-3}  
    \multicolumn{1}{|p{4cm}|}{ \text{Utilization TC0} \centering } &
    \multicolumn{1}{p{2.5cm}|}{ \text{78,06\%} \centering } &
    \multicolumn{1}{p{2.5cm}|}{ \text{86,94\%} \centering }
  \\  
  \cline{1-1}\cline{2-2}\cline{3-3}  
    \multicolumn{1}{|p{4cm}|}{ \text{Utilization TC1} \centering } &
    \multicolumn{1}{p{2.5cm}|}{ \text{89,41\%} \centering } &
    \multicolumn{1}{p{2.5cm}|}{ \text{87,01\%} \centering }
  \\  
  \cline{1-1}\cline{2-2}\cline{3-3}  
    \multicolumn{1}{|p{4cm}|}{ \text{Utilization TC2}\centering } &
    \multicolumn{1}{p{2.5cm}|}{ \text{93,10\% } \centering } &
    \multicolumn{1}{p{2.5cm}|}{ \text{86,92\%} \centering }
  \\  
    \cline{1-1}\cline{2-2}\cline{3-3}  
    \multicolumn{1}{|p{4cm}|}{ \text{Mean utilization}\centering } &
    \multicolumn{1}{p{2.5cm}|}{ \text{86,86\%} \centering } &
    \multicolumn{1}{p{2.5cm}|}{ \text{86,96\%} \centering }
  \\  
    \cline{1-1}\cline{2-2}\cline{3-3}  
    \multicolumn{1}{|p{4cm}|}{ \text{Block rate TC0}\centering } &
    \multicolumn{1}{p{2.5cm}|}{ \text{8,72\%} \centering } &
    \multicolumn{1}{p{2.5cm}|}{ \text{10,68\%} \centering }
  \\  
    \cline{1-1}\cline{2-2}\cline{3-3}  
    \multicolumn{1}{|p{4cm}|}{ \text{Block rate TC1}\centering } &
    \multicolumn{1}{p{2.5cm}|}{ \text{4,75\%} \centering } &
    \multicolumn{1}{p{2.5cm}|}{ \text{10,61\%} \centering }
  \\  
    \cline{1-1}\cline{2-2}\cline{3-3}  
    \multicolumn{1}{|p{4cm}|}{ \text{Block rate TC2}\centering } &
    \multicolumn{1}{p{2.5cm}|}{ \text{3,36\%} \centering } &
    \multicolumn{1}{p{2.5cm}|}{ \text{10,69\%} \centering }
  \\  
    \cline{1-1}\cline{2-2}\cline{3-3}  
    \multicolumn{1}{|p{4cm}|}{ \text{Mean block rate}\centering } &
    \multicolumn{1}{p{2.5cm}|}{ \text{5,61\%} \centering } &
    \multicolumn{1}{p{2.5cm}|}{ \text{10,66\%} \centering }
  \\  
  \hline

 \end{tabular} }
\end{table*}

Table~\ref{tableII} shows that the NN-PCM and the FRFS perform with equivalent efficiency in terms of link utilization. On the other hand, the NN-PCM performs better for TC1 and TC2 and worse for TC0. TC0 worse performance reflects the fact that it has the lowest priority among all classes in the NN-PCM class configuration mapping, and, due to that, higher priority classes TC1 and TC2 benefit from the ATCS sharing behavior. In conclusion, we verify that, as far as the utilization performance parameter is concerned, the NN-PCM is approximately equivalent to FRFS in terms of the proportionally ITM requirement.

The NN-PCM presents a much better result in terms of blocking users. This result shows that it enforces a proportional policy since it gets to the same network result (additional LSPs established and equivalent link utilization performance) interfering as less as possible with users.

\subsection{ITM Requirement 5 Conformance Accomplishment - Exceptionality}

The ITM exceptionality requirement defines that the management practice may temporarily and exceptionally throttle bandwidth-greedy users.

As an example, during congestion, bandwidth-greedy users can be exceptionally throttled. In the NN-PCM specific case, when congestion occurs, the traffic classes stop sharing bandwidth, and any previously shared bandwidth among classes is reclaimed back to its configured class.

\begin{figure*} [ht]
\centering
\includegraphics[width=0.65\textwidth]{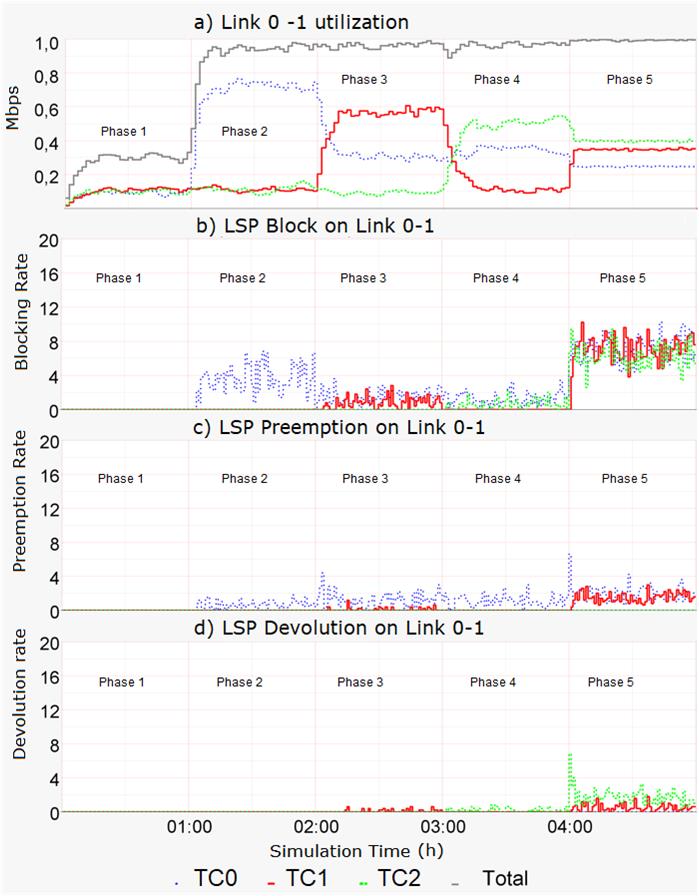} 
\caption{NN-PCM simulation with ATCS - Scenario 2 - ITM Exceptionality Conformance accomplishment}
\label{fig:Link0}
\end{figure*}

The simulation results for evaluating the ITM exceptionality for traffic scenario 2 is presented in Figure \ref{fig:Link0}. Let's first consider the NN-PCM behavior for all phases in sequence and then, identify the exceptional NN-PCM conformance.

Figure \ref{fig:Link0}.a illustrates the link utilization during the operation of the NN-PCM for simulation scenario 2. In phase 1 (low bandwidth demand), the blocking rate is null since there is an excess of available bandwidth for the link, and ATCS model shares the not used bandwidth among all users. In phases 2 to 4 (medium bandwidth demand), there is a high demand for classes 0, 1, 2 alternatively. In these phases, LSP block rate is higher due to the high occupation of the link, and the sharing among class is less frequent. In phase 5 (high bandwidth demand for all class - near congestion), the block rate is high for all TCs. This is because bandwidth demand is higher than its availability for all classes. The sharing stops, and each class tends to use its resources (Figure \ref{fig:Link0}.b).

LSP preemption and devolution rates have similar behavior, as illustrated in Figures \ref{fig:Link0}.c and \ref{fig:Link0}.d.

The exposed NN-PCM behavior for the link allows us to conclude that the trigger for applying the differentiated treatment among users is the occurrence of congestion, i. e., the LSP block, preemption, and devolution events occur only when bandwidth demand is beyond class or link availability. As such, the NN-PCM using the ATCS BAM assures not only an exceptional but also a temporary management practice, as required by ITM exceptionality.

During congestion, the NN-PCM has a non-discriminatory behavior. That is so because, during congestion, each class uses its predefined resources, and there is no discrimination among users belonging to the same class.

\subsection{NN-PCM Network Neutrality with ATCS - Results Summary}

In summary, the simulation results indicate as a proof-of-concept that the NN-PCM operation with ATCS accomplishes conformance by design to the BEREC network neutrality requirements, which are legitimate network management purpose, transparency, non-discrimination behavior, proportional behavior, and exceptional behavior.

Network neutrality is enforced inherently by the NN-PCM operation, with the network bandwidth being allocated to users grouped into traffic classes. The ATCS BAM model used in the NN-PCM manages the bandwidth allocation operation and the bandwidth sharing when traffic class resources become exhausted.

\subsection{ITM Requirement Conformance Accomplishing for Other BAM Models}

There are 3 basic BAM models (MAM, RDM, and ATCS) and several hybrids \cite{reale_g-bam:_2014}. This leads to the following question: Does the NN-PCM conformance result obtained for the ATCS model holds for other BAM models?

The non-discriminatory behavior (ITM requirement 3) of the NN-PCM holds for all models, as previously demonstrated (Section \ref{sec:non-discriminatory}).

The NN-PCM conformance to exceptionality requirement using other BAM model than ATCS can be intuitively derived from previous ATCS simulation results.  In effect, if the BAM model shares bandwidth, independently of the style and volume of bandwidth shared, it only blocks, preempts, or return resources when bandwidth demand is beyond class or link availability. Consequently, BAM behavior assures an exceptional and temporary management practice. The MAM model is a specific case in which there is no bandwidth sharing at all. In this case, there is no provision for exceptional or temporary practices, and users only dispute the available bandwidth per class.

The conformance with the proportionality requirement for other BAM models follows the same line of reasoning. In effect, bandwidth sharing results in outperforming the FRFS bandwidth allocation method. Consequently, other BAM models using distinct bandwidth sharing approaches do conform to proportionally ITM requirement. The MAM model without any sharing behavior is nearly equivalent to FRFS in terms of resource grant. In this case, eventual advantages or disadvantages concerning FRFS depend on the traffic input pattern, and, as such, we cannot guarantee that, specifically for MAM model, it will be fully compliant and always conform to ITM proportionality requirement.

In summary, all BAM models and hybrids, except for MAM, are compliant with the ITM requirements and do accomplish the BEREC network neutrality policy. The MAM model does not specifically conform to the proportionality ITM requirement 3.

\subsection{NN-PCM Quality of Service Related Considerations}

The current evaluation is a proof-of-concept that NN-PCM does achieve a neutral operation allocating bandwidth to users belonging to different traffic classes. The allocations remain neutral even if the bandwidth, as a resource, becomes scarce.

Looking from a different perspective, the neutral operation of the NN-PCM does not guarantee that all users in all classes will always get the bandwidth they want when they request it. That is so because resources (i.e., link capacity) are typically under-provisioned by network operators. Consequently, under traffic spike conditions, the bandwidth will probably not be available for all requesting users. Neutrality, as demonstrated, will be kept by the NN-PCM, but the NN-PCM operation can not necessarily maintain the quality of service (QoS) the users obtain in normal traffic conditions (non-congested). In this scenario, the lack of resources impacts all users, and the impact is inversely proportional to the priority they have (TC priority). Anyhow, from the QoS perspective, all users will experience a smaller or more significant impact. The NN-PCM effectively shares the scarcity of resources neutrally among users.

This work has not focused on evaluating the impact of bandwidth scarcity in the QoS obtained by users with a neutral operation. Future work will consider the analysis and evaluation of the quality of service parameters like packet loss and delay that can be effectively obtained for users with the network submitted to peaks of traffic or, equivalently, with the network deliberately using under-dimensioned links.

It is also important to highlight that multiple solutions can be achieved with the NN-PCM (BAM) that can keep an equilibrium between network neutrality and QoS impact. As an example, a simple and straightforward approach for the operator business model would be to have a highly prioritized traffic class that, with congestion, uses the extra bandwidth of low priority classes by design. In other words, it means that the network operator can define its classes of traffic in such a way that, for some users, QoS will be less impacted and in a previously simulated range. The NN-PCM operation based on a BAM allows this kind of previous simulated configuration. In brief, the NN-PCM neutral operation does not necessarily maintain QoS, but it allows countermeasures and the possibility to design a contingency solution previously.

\section{Final Considerations}\label{sec5}

The NN-PCM is an innovative network neutrality tool based on a BAM model. The NN-PCM network neutrality accomplishment of the European BEREC reasonable Internet Traffic Management (ITM) practices was modeled and validated through the analysis of BAM models behavior and simulations. The results demonstrate that NN-PCM using the ATCS BAM model is legitimate, transparent, non-discriminatory, proportional, and exceptional.

A relevant operational aspect of the NN-PCM is its capability to accomplish network neutrality by design. In effect, the conformance to the BEREC ITM policy results from the inherent BAM behavior associated with the grouping of users in classes. In other words, NN by design means that the network manager chooses the BAM, configure it, and, following that, the NN-PCM operation allocates bandwidth to users in conformance with the BEREC network neutrality policy. To the best of our knowledge, network neutrality by design is unexplored by other NN technical deployments that focus mainly on detecting NN violations, reacting to them, and discriminating traffic.

Transparency concerning the business model adopted by network management is another positive aspect achievable by the NN-PCM. The business model in this paper corresponds to the allocation of users in classes and the definition of how much bandwidth is allocated to each class. The NN-PCM allows users to know two pieces of information about the business model. Firstly, it is possible to advertise to users the amount of resources they collectively have that technically corresponds to the minimum bandwidth constraint per class. Secondly, the users may know how the network will manipulate the available bandwidth per class and among classes. This second point corresponds to the configured BAM model operation and behavior. Whenever this information is made public, the NN-PCM accomplishes transparency by exposing to users what they have as resources and how these resources are allocated among them. By exposing the BAM configuration information, the NN-PCM contributes to solving a current network neutrality deployment problem, which is why most NN deployments have opaque policies that are hard to perceive or not perceived at all by the users.

The focus of the NN-PCM development is to demonstrate that, whether the available bandwidth configured for each class is under or over-dimensioned, the allocation of bandwidth to the user remains neutral and respects the BEREC requirements for network neutrality. This result was achieved, and, in summary, it means that the NN-PCM does accomplish BEREC network neutrality conformance with a neutral allocation of bandwidth among all users, regardless of their classes.

\bibliography{wileyNJD-AMA}%

\clearpage

\section*{Authors Biography}

\begin{biography}{\includegraphics[width=66pt,height=86pt]{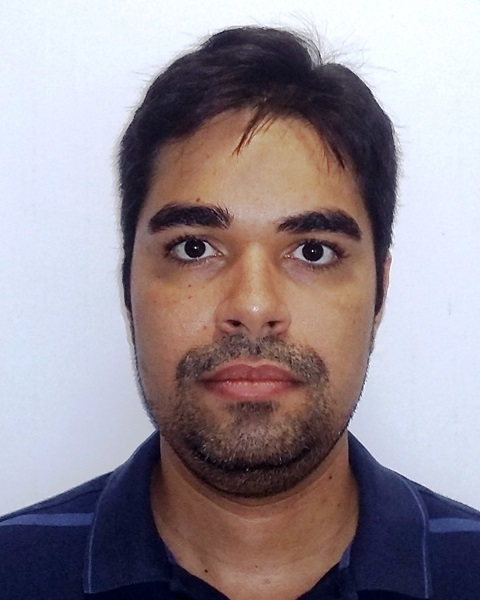}}{\textbf{M. Sc. David S. S. Barreto.} Bachelor in Electrical Engineering by Faculdade AREA1 and Masters Degree in Computer Science at Salvador University (UNIFACS). David is a member of IPQoS Research Group and RePAF Project (Resource Allocation Framework for MPLS, Elastic Optical Networks and IoT) at UNIFACS and currently works at ANATEL (Brazilian Telecommunications Regulatory Agency). His research interests include Network Neutrality, Coverage and Quality measurements in mobile networks, Performance, QoS and QoE.}
\end{biography}
\hfill\break

\begin{biography}{\includegraphics[width=66pt,height=86pt]{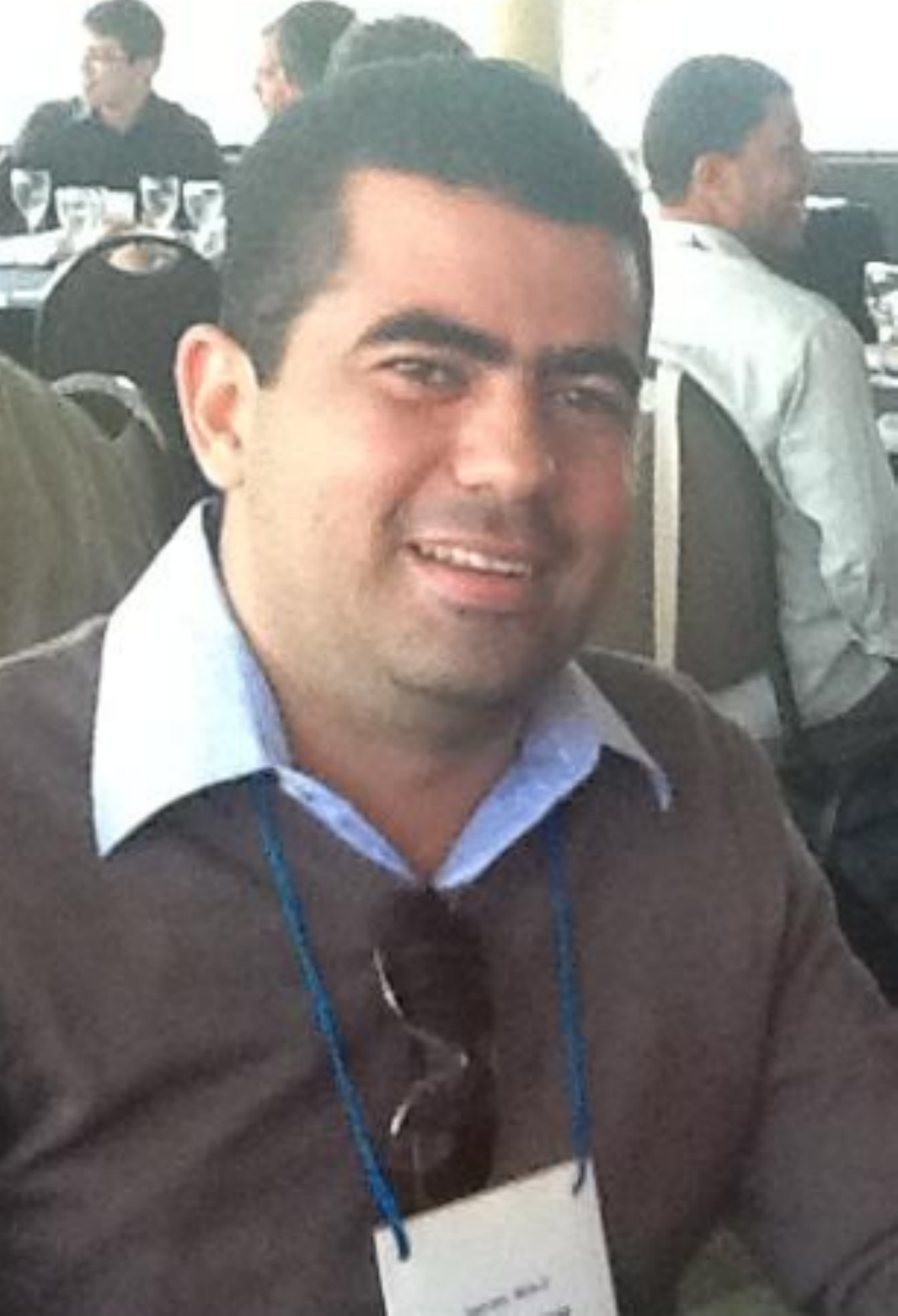}}{\textbf{Prof. Dr. Rafael F. Reale.} Professor at Institute Federal of Bahia (IFBA). Ph.D. in Computer Science with DMCC (UFBA/UNIFACS/UEFS), MSc. in Computer Systems by Salvador University - UNIFACS (2011) and bachelor in Informatics by Universidade Católica do Salvador (2005). Professor at Instituto Federal da Bahia (IFBA). His current research interests include Bandwidth Allocation Model, MPLS, DS-TE, Autonomy, QoS, Future Internet Architectures, Software-Defined Networks and Cognitive Networks.}
\end{biography}

\hfill\break

\begin{biography}{\includegraphics[width=66pt,height=86pt]{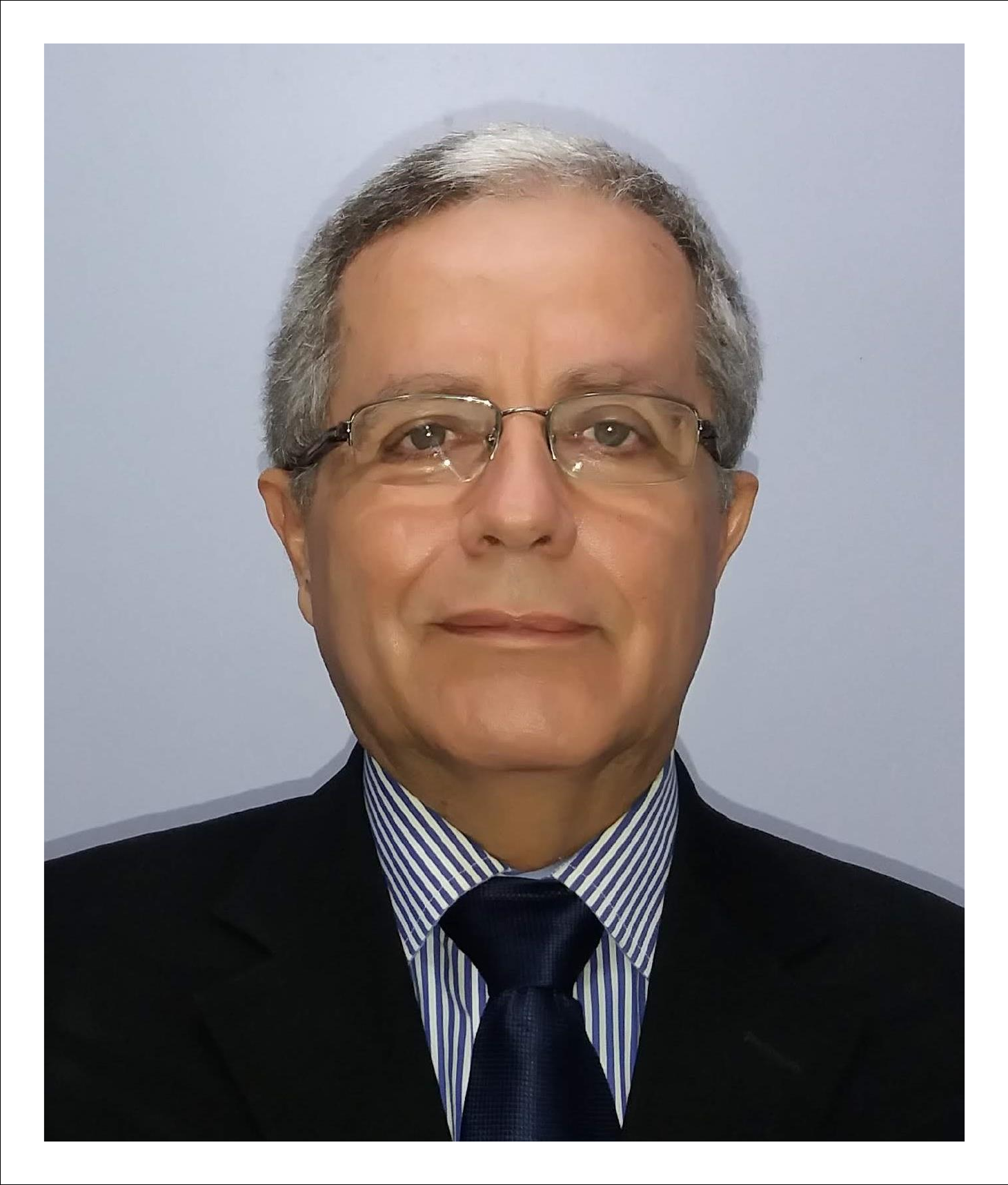}}{\textbf{Prof. Dr. Joberto S. B. Martins.} Ph.D. in Computer Science at Université Pierre et Marie Curie - UPMC, Paris (1986), PosDoc at ICSI/ Berkeley University (1995), and PosDoc Senior Researcher at Paris Saclay University - France (2016). International Professor at Hochschule für Technik und Wirtschaft des Saarlandes - HTW (Germany) (since 2004) and Université d'Evry (France). Full Professor at Salvador University (UNIFACS) on Computer Science, Director of NUPERC and IPQoS research groups with research interests on Cognitive Management, Resource Allocation, Machine Learning, SDN/ OpenFlow, Internet of Things, Smart Grid and Smart Cities.}
\end{biography}
\hfill\break

\end{document}